# Ab initio calculation of electronic band structure of $Cd_{1-x}Fe_xSe$

Matanat A. Mehrabova[*a,b], Elshad A. Allahyarov[c,d],

Niyazi H. Hasanov[e], Nurana R. Gasimova[f]

[a]*Azerbaijan Technical University, Baku, Azerbaijan*
[b]*Institute of Radiation Problems, Ministry of Science and Education, Baku, Azerbaijan*
[c]*Heinrich-Heine University Dusseldorf, Dusseldorf, Germany*
[d]*Macromolecular Science, CWRU Cleveland, USA*
[e]*Baku State University, Baku, Azerbaijan*
[f]*Azerbaijan University of Architecture and Construction, Baku, Azerbaijan*

**Abstract**

The purpose of this work was to calculate the electronic band structure of $Cd_{1−x}Fe_xSe$. Ab initio, calculations are performed in the Atomistix Toolkit program within the Density Functional Theory and Local Spin Density Approximation on Tight Tiger basis. We have used Hubbard U potential $U_{Fe}$ = 2.42eV for 3d states for Fe ions. Supercells of 8 and 64 atoms were constructed. After the construction of $Cd_{1−x}Fe_xSe$ (*x*=6.25%; 25%) supercells, atom relaxation and optimization of the crystal structure were carried out. Electronic band structure, and density of states were calculated, and total energy have been defined in antiferromagnetic and ferromagnetic phases. The band gap for the $Cd_{1-x}Fe_xSe$, *x*=0.06 in ferromagnetic phase is equal to $E_g$= 1.77 eV, in antiferromagnetic phase $E_g$=1.78 eV. For *x*=0.25 $E_g$= 1.92 eV. Antiferromagnetic phase considered more stable. Our calculations show that the band gap increases with the increases in Fe ion concentration.

## 1. Introduction

Iron-containing II-VI compounds are a new class of semimagnetic semiconductors (SMSC) compared to well-known manganese-based SMSC. The data on Fe-based SMSC are far less than those on Mn-based SMSC. In these SMSC, the 3d shell of $Fe^{2+}$ ions contains one more electron than the half-filled 3d shell of $Mn^{2+}$. The solubility of iron in II-VI compounds is relatively smaller than that of manganese: about 2 to 5 at.% in the tellurides and up to 20 at.% in the selenides. Off-equilibrium methods, like molecular beam epitaxy, can introduce larger concentrations [1,2].

---

[*]*Corresponding author*. Tel.: +994-50-731-81-77
*E-mail address*: metanet.mehrabova@aztu.edu.az; *ORCID ID*: 0000-0003-4417-0522.

The $Cd_{1-x}Fe_xSe$ thin films are of great interest, for their wide-range application in semiconducting devices, such as photovoltaic, optoelectronic devices, radiation detectors, laser materials, thermoelectric devices, solar energy converters, video on devices, etc. [3,4].

$Cd_{1-x}Fe_xSe$ has a sphalerite structure and crystallize in the wurtzite structure. The $Cd_{1-x}Fe_xSe$ solid solutions are single-phase for *x* < 0.15.

The band gap of $Cd_{1-x}Fe_xSe$ SMSC can be varied by means of the composition between 1.44 eV for FeSe to 1.72 eV for CdSe, almost covering the entire visible range. However, for efficient optoelectronic device applications, the material is usually between CdSe in which very high sensitivity is possible but response time is high, and CdFeSe in which a lower response time is possible at the cost of some loss in sensitivity.

In our previous papers the synthesis of solid solutions and thin film grow processes of SMSC including $Cd_{1-x}Fe_xSe$ were studied, their electrical and optical properties were investigated [5,6]. In the present paper theoretically investigated the electron band structure of $Cd_{1-x}Fe_xSe$ by density functional theory (DFT) DFT theory using ab initio calculations.

## 2. Results and discussions

The calculations are based on the first-principles pseudopotential method within the DFT and the local spin density approximation (LSDA) using the Tight Tier basis set. All calculations are performed by using Atomistix Toolkit (ATK) program [7].

$Cd_{1-x}Fe_xSe$ supercell of 8 and 64 atoms were constructed. It was used Hubbard *U* potential $U_{Fe}$ = 2.42 eV for 3d states for $Fe^{2+}$ ions [8, 9]. After the construction of $Cd_{1-x}Fe_xSe$ (*x* = 0; 6.25 %; 25 %) supercells, atom relaxation and optimization of the crystal structure were carried out to eliminate forces and minimize stresses. The equilibrium lattice parameters have been computed by minimizing the crystal total energy calculated for different values of lattice constant. Firstly electron band structure (EBS) was calculated for CdSe semiconductor compound (Figure 1). The calculated band gap was 1.73 eV.

The electron band structure of $Cd_{1-x}Fe_xSe$ SMSC is determined from the projected density of states (PDOS). The analysis of these graphs shows that in the valence band, the electron band structure of $Cd_{1-x}Fe_xSe$ consists of three parts: (1) the upper part of the valence band is mainly formed by *p*-orbitals of Se and Cd atoms, *s*-orbitals of Cd and Fe atoms with some contribution of *d*-orbitals of Fe atoms; (2) the middle part is formed by *d*-orbitals of Cd atoms, which are 7 eV lower than the valence band maximum (3) the lower part is formed by *s*-orbitals of Se and Fe atoms, and *p*-orbitals of Fe atoms which are located 18 eV lower than the valence band maxim.

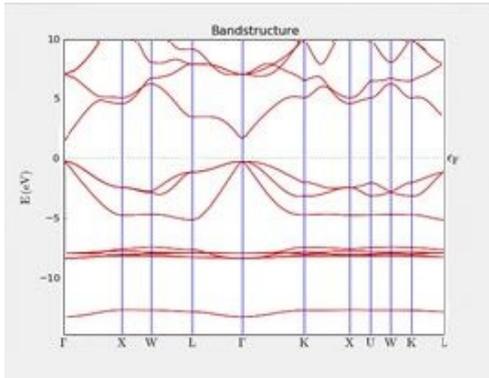

**Fig. 1.** Band structure of CdSe.

The bottom of the conductivity band is formed by *s*-and *p*-orbitals of Fe atoms and *p*-orbitals of Cd atoms, *d*-orbitals of Se atoms.

EBS and density of states (DOS) for $Cd_{1-x}Fe_xSe$, $x=0.25$ were calculated, and band gap, the total energy have been defined as $E_g$= 1.92 eV, $E_t$=-6308.42730 eV respectively. In Figure 2 is given the bulk configuration and band structure of $Cd_{1-x}Fe_xSe$, $x=0.25$.

EBS and density of states (DOS) for $Cd_{1-x}Fe_xSe$, $x=0.06$ were calculated and the band gap, DOS, total energy has been defined in antiferromagnetic (AFM) and ferromagnetic (FM) phases (Figure 3).

The band gap for the $Cd_{1-x}Fe_xSe$, $x=0.06$ in ferromagnetic phase is equal to $E_g$ = 1.77 eV and total energy is equal to $E_t$ = -60818.04032 eV (Figure 4).

The band gap for the $Cd_{1-x}Fe_xSe$, $x=0.06$ in antiferromagnetic phase is equal to $E_g$ = 1.78 eV and total energy is equal to $E_t$ = -60818.03595 eV (Figure 5).

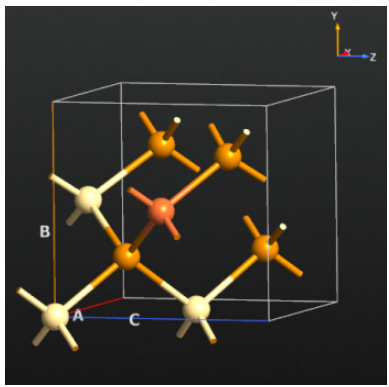
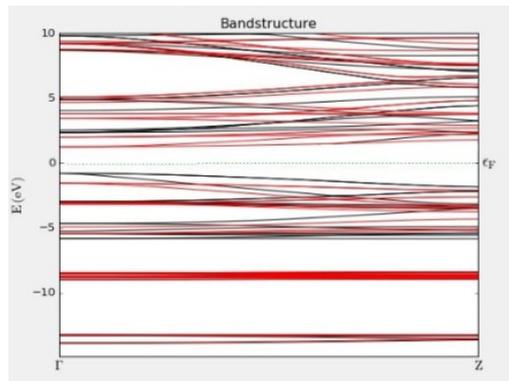

a)  b)

**Fig. 2.** $Cd_{1-x}Fe_xSe$, $x=0.25$ a) bulk configuration, b) electron band structure

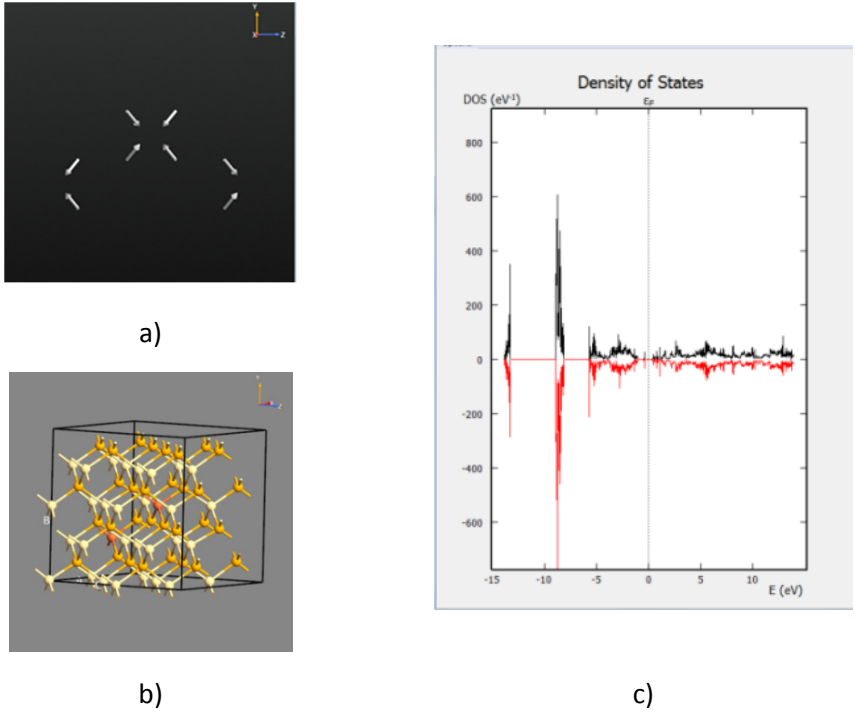

**Fig. 3.** $Cd_{1-x}Fe_xSe$, *x*=0.06: a) forces; b) bulk configuration; c) DOS

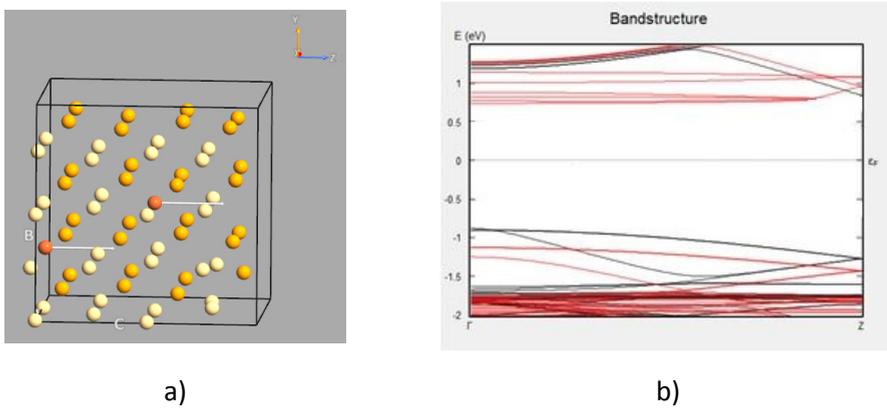

**Fig. 4.** $Cd_{1-x}Fe_xSe$, x=0.06, ferromagnetic phase:
a) bulk configuration; b) electron band structure

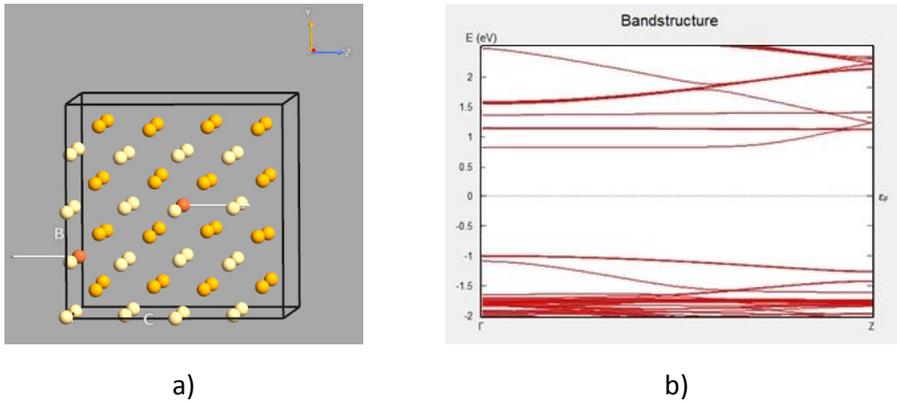

a)            b)

**Fig. 5.** $Cd_{1-x}Fe_xSe$, $x$=0.06, antiferromagnetic phase: a) bulk configuration; b) electron band structure

The calculated band gap much closer to literature data [10,11]. It was defined that total energy in AFM phase more than in FM phase $E_t$ (AFM) › $E_t$ (FM), therefore AFM phase considered more stable. Band gap in AFM more than in FM phase $E_g$(AFM) › $E_g$ (FM).

## 3. Conclusion

It was ab initio calculated band structure of $Cd_{1–x}Fe_xSe$ SMSC by using of Atomistix Toolkit program within the Density Functional Theory and Local Spin Density Approximation on Tight Tiger basis. Hubbard $U$ potential $U_{Fe}$ = 2.42eV was used for 3d states for $Fe^{2+}$ ions. Supercells of 8 and 64 atoms were constructed. After the construction of $Cd_{1–x}Fe_xSe$ ($x$=6.25%; 25%) supercells, atom relaxation and optimization of the crystal structure were carried out. Electronic band structure, and density of states were calculated, and total energy have been defined in antiferromagnetic and ferromagnetic phases. Our calculations show that the band gap increases with the increases in $Fe^{2+}$ ion concentration.